%% file: complexity_of_ddproblem.tex
\documentclass{article}
\input{lib_qwak.tex}

\title{Oszacowanie złożoności problemu rozgrywki w otwarte karty w brydżu}
\author{Piotr Beling \\ Uniwersytet Łódzki, Wydział Matematyki i Informatyki}

\begin{document}
\maketitle

\begin{abstract}
Artykuł zawiera analizę złożoności problemu rozgrywki w~otwarte karty w brydżu, przy użyciu miar zaproponowanych przez Louisa Victora Allisa w~\cite{complexity:Allis94}. 
Oszacowane są w nim złożoności przestrzeni stanów i~drzewa wspomnianej gry.
\end{abstract}


\tableofcontents

\section{Miary złożoności gier}

Louis Victor Allis w~\cite{complexity:Allis94} zaproponował i, posługując się przykładami wielu gier, pokazał zasadność następujących miar oceny trudności znalezienia rozwiązania dwuosobowych gier z~doskonałą informacją, o~sumie stałej\footnote{Więcej na temat takich gier i metod ich rozwiązywania można znaleźć w bardzo licznej literaturze, np. \cite{Plaat96/mtd, BelingWarcaby, qian, chesswiki, sgp}.}: 
\begin{itemize}
  \item Złożoność przestrzeni stanów (ang. state-space/search-space complexity) jest to liczba legalnych, różnych pozycji osiągalnych z~pozycji początkowej. 
  \item Złożoność drzewa gry (ang. game tree complexity) jest to liczba liści w~drzewie poszukiwań rozwiązania początkowej pozycji gry.
  
Gdy przez $M$ oznaczymy drzewo poszukiwań algorytmu Minimax o~korzeniu w~$J$, to drzewem poszukiwań rozwiązania węzła $J$ (ang. solution search tree of a~node $J$) nazwiemy poddrzewo $M$ składające się ze~wszystkich węzłów $M$ aż do~głębokości rozwiązania węzła $J$.

Głębokość rozwiązania węzła $J$ (ang. solution depth of~a~node $J$) to~minimalna głębokość na~jaką trzeba znać $M$ aby móc ustalić ile wynosi wartość minimaksowa pozycji $J$.
\end{itemize}

\begin{table}
\centering
\begin{tabularx}{\textwidth}{Xp{2.9cm}p{1.8cm}p{1.0cm}}
\hline
gra & \multicolumn{2}{l}{logarytm dziesiętny ze złożoności} & źródła\\
	& przestrzeni stanów & drzewa gry & \\
\hline
Kółko i krzyżyk (3x3) & 3 & 5 & \cite{complexity:Allis94} \\
Czwórki (Connect Four, 7x6) & 13 & 21 & \cite{complexity:Allis94, allis:connect4} \\
Rozgrywka w~widne w~brydżu (średnia dla 13 lew) & 17 & 29 & \\
Warcaby angielskie (amerykańskie) & 18 & 31 & \cite{complexity:Allis94} \\
Othello (Reversi, 8x8) & 28 & 58 & \cite{complexity:Allis94} \\
Warcaby międzynarodowe (10x10) & 30 & 54 & \cite{complexity:Allis94} \\
Szachy & 47 & 123 & \cite{complexity:Allis94, complexity:chess} \\
Hex (11x11) & 57 & 98 & \cite{complexity:Herik} \\
Gomoku (15x15, freestyle) & 105 & 70 & \cite{complexity:Allis94} \\
Go (19x19) & 171 & 360 & \cite{complexity:Allis94, complexity:GO} \\
\hline
\end{tabularx}
\caption{Szacowana złożoność wybranych gier.}
\label{tab:complexity}
\end{table}

Wyżej opisane miary są~dość proste i~niezbyt dokładne. Nie uwzględniają one wielu czynników (np. istnienia symetrii, możliwości budowania bazy końcówek, itd.).
W~pracy \cite{complexity:Heule} można znaleźć analizę dotychczas rozwiązanych gier która wskazuje, że~zbiór tych gier różni się nieco od~zbioru gier które powinny być~rozwiązane wg.~miar zaproponowanych przez Allisa.
Pomimo to, miary te~wciąż są~bardzo popularne. Prawdopodobną przyczyną tego stanu rzeczy jest relatywna łatwości oszacowania ich~wartości dla~poszczególnych gier. Opracowania \cite{complexity:Allis94, complexity:Herik, complexity:Heule, wikipedia:game_complexity} zawierają szacowane wartości zdefiniowanych miar dla~wielu popularnych gier logicznych. Część z~nich można znaleźć w~tabeli \ref{tab:complexity}.

Dokładniejszą analizę i~dodatkowe propozycje związane z~szacowaniem trudności gier zawierają m.in. prace \cite{complexity:Herik} i~\cite{complexity:Heule}.
Autorzy artykułu \cite{complexity:Herik}, zwracają m.in. uwagę, że w~grach w~których podczas trwania rozgrywki maleje liczba dostępnych stanów gry na~ogół możliwe jest zbudowanie bazy końcówek\footnote{Baza końcówek\index{baza!końcówek} zawiera pozycje które mogą pojawić się w~końcowej fazie gry, ich wartości minimaksowe i, opcjonalnie, najlepsze posunięcie jakie można w~nich wykonać lub liczbę ruchów do końca gry. Więcej informacji i~algorytm budowania bazy końcówek można znaleźć np. w~\cite{BelingWarcaby}.} (i~zapamiętanie jej przy użyciu akceptowalnie małej przestrzeni dyskowej), co~znacznie ułatwia znalezienie rozwiązania.
Przykładowo, przy intensywnym wykorzystaniu bazy końcówek, udało się znaleźć rozwiązanie dla~tak złożonej gry jak warcaby amerykańskie\cite{schaeffer:checkers}.
Grami spełniającymi wspomniany warunek są~także poszczególne problemy rozgrywkowe w~otwarte karty. Niestety, na~ogół zapisanie obszernej bazy końcówek dla~wszystkich elementów z~rodziny takich problemów (różniących się początkowym rozłożeniem kart) nie~jest możliwe.

\section{Problemu rozgrywki w otwarte karty (i oznaczenia)}
Problem rozgrywki w~otwarte karty w~brydżu polega na~wyznaczeniu optymalnej linii rozgrywki przy założeniu pełnej wiedzy o~rozkładzie (wszystkich) kart i~bezbłędnej gry wszystkich uczestników.

Problem ten i~jego efektywne rozwiązanie są~bardzo ważne, zarówno z~punktu widzenia (statystycznej) analizy rozdań przez brydżystów (por. np. \cite{ginsberg/lawOfTotalTricks}) jaki i~jako część składowa algorytmów wyznaczających sub\nobreakdash-optymalne decyzję przy kartach zakrytych (zarówno w~rozgrywce jak i licytacji czy wiście -- por. m.in. \cite{ginsberg/how_comp_play_bridge, ginsberg/expert_level_prog__bridge_short, ginsberg/imperf_inf_in_game__bridge_full, StateOfBridge, Levy1989}). Przykładowo, w~rozgrywce, można rozwiązać szereg problemów rozgrywki w~otwarte karty, dla wielu układów kart ,,pasujących'' do~znanych kart, licytacji i~wcześniejszej gry przeciwników i~zagrać kartę która okazała się wygrywająca w~największej liczbie przypadków. Proszę zauważyć, że takie postępowanie wymaga, by problemy składowe będące grą w~otwarte karty rozwiązywać bardzo szybko. Dlatego wiele osób, włożyło mnóstwo pracy w~opracowanie technik to~umożliwiających (np. \cite{beling/partition_search, beling/phd, beling/mis2011, DDS/Berlekamp1962, DDS/Berlekamp1963, Chang/building_fast_dds, Mossakowski2009}).

Kolejne rozdziały zawierają oszacowanie wartości złożoności przestrzeni stanów oraz drzewa gry dla problemu rozgrywki w~otwarte karty. Będę w nich posługiwał się następującymi oznaczeniami charakteryzującymi rozpatrywaną grę:
\begin{itemize}
  \item $\S$ -- przestrzeń stanowa (zbiór wszystkich pozycji),
  \item $\N: \S \to 2^\S$ --- funkcja wyznaczająca następniki danej pozycji, tj. $s' \in \N(s)$ oznacza, że $s'$ jest osiągalny z~$s$ poprzez wykonanie jednego ruchu;
  \item $\SUITS$ -- zbiór kolorów (typowo: trefle, kara, kiery, piki);
  \item $\HANDS$ -- zbiór rąk (w brydżu: North, East, South, West);
  \item $R = |\HANDS|$ --- liczba rąk (w~brydżu: $R=4$);
  \item $K$ --- liczba kart w~posiadaniu każdej z~rąk na~początku gry (w~brydżu: $K=13$).
\end{itemize}

\section{Złożoność przestrzeni stanów}

Wielkość przestrzeni stanowej ($|\S|$) i~tym samym jej złożoność, dla problemu rozgrywki w~otwarte karty, można oszacować z~góry w~następujący sposób:

Liczbę rozłożeń kart, takich że~jedna z~rąk posiada $k$ kart (gdzie $0 < k \leq K$), zaś każda z~pozostałych ma~ich $k$ albo $k-1$ (w~tym drugim przypadku, jedną kartę dołożyła do bieżącej lewy) można oszacować z~góry przez:
\begin{equation}\label{complexity:f} f(k)=\binom{K}{k}^R (1 + R \sum_{h=1}^{R-1}k^h) \end{equation}
Pierwszy czynnik iloczynu we~wzorze \eqref{complexity:f} wiąże się z~liczbą $k$\nobreakdash-elementowych podzbiorów kart każdego z~graczy.
Drugi szacuje liczbę możliwych wyborów kart (spośród wspomnianych $k$\nobreakdash-elementowych podzbiorów) wchodzących w~skład tworzonej lewy.
Gdy nie ma kart w~lewie, jest jeden możliwy wybór. Gdy jest $h$ kart w~lewie (gdzie $h \in \{1, \ldots, R-1\}$), każda spośród co najwyżej $R$ rąk może lewę rozpocząć, zaś z~każdej spośród $h$ rąk które dołożyły do lewy mogła być zagrana jedna spośród co najwyżej $k$ kart.

Oznaczę $f(0)=1$ (liczba rozłożeń pustego zbioru kart).

Liczbę stanów w~których jedna z~rąk posiada $k$ kart, zaś każda z~pozostałych ma~ich $k$ albo $k-1$ jest nie większa niż:
\begin{equation}\label{complexity:fp} f_p(k) = (K-k+1)f(k) \end{equation}
Czynnik $(K-k+1)$ związany jest z~lewami wcześniej zdobytymi przez poszczególne pary. Po zagraniu $K-k$ pełnych lew, każda z~nich mógła, potencjalnie, zgromadzić od~$0$ do $K-k$ lew, co~daje $K-k+1$ możliwości.

Ostatecznie, liczba wszystkich stanów można oszacować, z~góry, następująco:
\begin{equation} |\S| \leq \sum_{k=0}^{K} f_p(k) \end{equation}
Dla~brydża ($R=4$, $K=13$) szacowanie to~wynosi około $2 \cdot 10^{17}$.

Ponieważ przy obliczeniach często można nie rozróżniać pozycji różniących się jedynie liczbą wcześniej wziętych lew, to dla oceny złożoności problemu, ważna jest też wartość:
\begin{equation} \sum_{k=0}^{K} f(k) \end{equation}
która dla~brydża ($R=4$, $K=13$) jest liczbą równą około $3 \cdot 10^{16}$.


\section{Złożoności drzewa gry}

Ponieważ rozgrywka w~otwarte karty ma~stałą długość (zawsze kończy się po~zagraniu wszystkich $R \cdot K$, w brydżu 52, kart), to złożoność drzewa gry jest równa wielkość drzewa gry (ang. game tree size) zdefiniowanej jako liczba wszystkich możliwych sposobów rozegrania gry\cite{wikipedia:game_complexity}, lub równoważnie jako liczba liści w~drzewie minimaksowym o~korzeniu w~pozycji początkowej $s_I$.

W zależności od~początkowego układu kart wielkość ta, może wynieść nawet:
\begin{equation}\label{complexity:maxtreesize} K!^{R} \end{equation}
Taka wielkość drzewa gry wystąpi w~rozdaniach w~których każdy z~graczy, będzie mógł za każdym razem zagrać dowolną z~posiadanych kart.
Przykładowo: wistujący będzie posiadał wszystkie karty w~jednym kolorze i~będzie to~kolor atutowy lub gra będzie w bezatu (wszyscy dokładający do kolejnych lew nie będą mieli żadnej karty w kolorze wistu). 

Dla brydża liczba opisana wzorem \eqref{complexity:maxtreesize} wynosi około $1{,}5 \cdot 10^{39}$.

Dolne ograniczenie na~wielkość drzewa gry, można uzyskać np. zakładając, że~każdy z~dokładających do~lewy graczy będzie miał tylko jedną możliwość:
\begin{equation}\label{complexity:mintreesize:weak} K! \end{equation}
Dla brydża liczba ta jest równa $13! \approx 6{,}2 \cdot 10^9$.

Następujące, dokładniejsze szacowanie dolnego ograniczenia wielkości drzewa problemu rozgrywki w otwarte karty w brydżu, zawarł w~swojej rozprawie doktorskiej \cite{frank:Search_and_Planning_under_Incomplete_Information_96} Ian Frank:
Niech $s_{rk}$ będzie liczbą kart posiadaną przez $r$\nobreakdash-tą rękę w $k$\nobreakdash-tym kolorze (dla dowolnej ręki $r \in \HANDS$ i koloru $k \in \SUITS$).
Naturalnie $\sum_{k \in \SUITS} s_{rk} = K$.
Najmniejsza liczba możliwych do~wyboru, w~trakcie całego rozdania, zagrań z~ręki $r$ związana jest z~takim przebiegiem gry, że $r$ nigdy nie~będzie wistowała i~równocześnie zawsze będzie dokładała do~koloru. Liczba ta~wynosi:
\begin{equation}\label{complexity:mintreesize:player:strong} \prod_{k \in \SUITS} s_{rk}! \end{equation}
Stąd, liczba możliwych, różnych przebiegów całej gry nie może być mniejsza niż:
\begin{equation}\label{complexity:mintreesize:strong} \prod_{r \in \HANDS} \prod_{k \in \SUITS} s_{rk}! \end{equation}
Liczba opisana wzorem \eqref{complexity:mintreesize:strong} jest najmniejsza dla rozdań w~których wszystkie ręce będą możliwie najbardziej zrównoważone. W~brydżu jest ona najmniejsza gdy każda z~rąk ma układ 4\nobreakdash-3\nobreakdash-3\nobreakdash-3 i~wtedy, wynosi ona~około $7{,}2 \cdot 10^{14}$. 
Ian Frank policzył także, że dla~losowego rozdania, oczekiwana wartość dolnego ograniczenia wielkości drzewa gry opisanego wzorem \eqref{complexity:mintreesize:strong} wynosi dla~brydża $1{,}05 \cdot 10^{18}$.

\begin{figure}
\begin{tikzpicture}
 \begin{axis}[
 	width=\textwidth,
 	height=0.8\textheight,
  	xbar=2pt,	
  	bar width=\textheight/50,
 	ytick=data,
 	nodes near coords, nodes near coords align={horizontal},	
 	xmin=0,
 	ylabel=Numer lewy,
 	xlabel=Średnia liczba możliwych do~dołożenia kart,
 	y dir=reverse,
 	legend style={
		legend pos=south east
	}
 ]
  \addplot table[x=moves,y=trick] {
  		trick moves		
		1 3.28053
		2 3.31925
		3 3.35428
		4 3.38003
		5 3.38596
		6 3.36102
		7 3.29322
		8 3.16902
		9 2.97278
		10 2.68588
		11 2.28497
		12 1.73835
		13 1.00
	};
	\addlegendentry{gra w bez atu}

  \addplot table[x=moves,y=trick] {
  		trick moves				
		1 3.28053
		2 3.30582
		3 3.33294
		4 3.35223
		5 3.35367
		6 3.32641
		7 3.25848
		8 3.13636
		9 2.94431
		10 2.66348
		11 2.26999
		12 1.73138
		13 1.00
	};
	\addlegendentry{gra atutowa}
 \end{axis}
\end{tikzpicture}
\caption{Średnia liczba kart możliwa do dołożenia przez graczy do~kolejnych lew w brydżu. Wartości zostały obliczone doświadczalnie, przy założeniu losowych przebiegów rozgrywek.}
\label{plot:avg_moves_at_trick}
\end{figure}
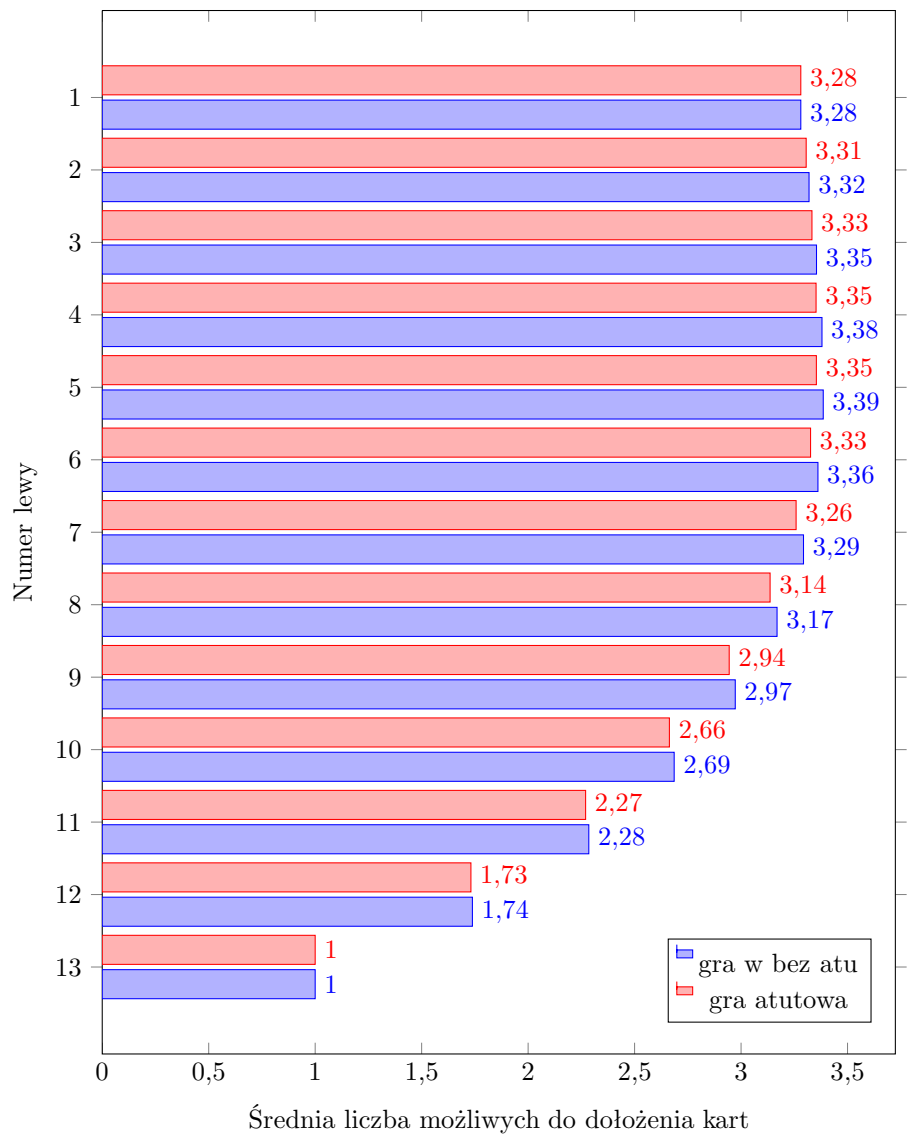

\begin{table}
\centering
\begin{tabularx}{\textwidth}{X|c|c}
\multicolumn{1}{r|}{Typ gry:}	& bezatutowa				& w~kolor	\\	
\hline
Średnia szacowana wielkość drzewa gry			& $2{,}4 \cdot 10^{29}$		& $6{,}8 \cdot 10^{28}$		\\
Średni błąd kwadratowy wartości średniej		& $0{,}2 \cdot 10^{29}$		& $0{,}6 \cdot 10^{28}$		\\
Odchylenie standardowe							& $7{,}4 \cdot 10^{33}$		& $3{,}5 \cdot 10^{33}$		\\
Minimalna wartość w~próbie	& $8{,}9 \cdot 10^{18}$	& $8{,}9 \cdot 10^{18}$		\\
Maksymalna wartość w~próbie	& $1{,}5 \cdot 10^{39}$	& $1{,}5 \cdot 10^{39}$		\\	
Wielkość próby									& $4 \cdot 10^{11}$			& $4 \cdot 10^{11}$		\\
\end{tabularx}
\caption{Znaleziona numerycznie, szacowana wielkość drzewa gry.}
\label{tab:dd_tree_size}
\end{table}

Rozpiętość pomiędzy dolną ($7{,}2 \cdot 10^{14}$), a górną ($1{,}5 \cdot 10^{39}$) szacowaną wielkością drzewa gry dla problemu rozgrywki w~otwarte karty w~brydżu jest bardzo duża. Prawdopodobnie, znaczne są też różnice wielkości samych drzew uzyskanych dla różnych, początkowych układów kart.

Jaka jest jednak oczekiwana wielkością drzewa dla tej gry? I jaki jest jego oczekiwany kształt? Wiadomo, że~co czwarta pozycja jest wistowa. Z~ręki wistującej można zagrać dowolną z~posiadanych kart i w~związku z~tym stopnie rozgałęzienia kolejnych pozycji wistowych wynoszą odpowiednio $13, 12, \ldots, 1$. Jaka jest jednak średnia liczba kart możliwa do dołożenia przez graczy do~kolejnych lew?

By~oszacować jaka jest odpowiedź na~wyżej postawione pytania, wykonano następujący eksperyment (niezależnie dla gier atutowych i w~bez atu):
\begin{enumerate}
	\item Wylosowano próbę $P$ rozdań (gier).
	
	\item Dla każdego z~wylosowanych rozdań $p \in P$ przeprowadzono losową rozgrywkę (kolejne zagrania były wybierane z~jednostajnym prawdopodobieństwem) uzyskując ciąg pozycji $s_0(p), \ldots, s_{51}(p)$ określających jej przebieg.
	
	\item Oszacowano liczbę kart możliwą do~dołożenia przez graczy do~$n$\nobreakdash-tej lewy ($n = 1, \ldots, 13$) w~grze $p \in P$ jako:
	 \[\sym{deg}_p(n)=\sum_{i=1}^{3} |\N(s_{4(n-1)+i}(p))| / 3\]
	 Uzyskane dla~kolejnych rozdań z~próby $P$ wyniki uśredniono:
	 \[\sym{deg}(n)=\sum_{p \in P} \sym{deg}(n) / |P|\]
	 Rezultat przedstawiono na wykresie \ref{plot:avg_moves_at_trick}.
	 
	\item Oszacowano wielkość drzewa gry $p \in P$ jako: $\prod_{i=0}^{51} |\N(s_i(p))|$. Wielkość tą uśredniono dla całej próby $P$, wyniki zebrano w tabeli \ref{tab:dd_tree_size}. 
\end{enumerate}

Wykres \ref{plot:avg_moves_at_trick} przedstawia średnią liczba kart możliwą do dołożenia przez graczy do~kolejnych lew w brydżu.
Nieco większe wartości uzyskane dla rozgrywek w~bez atu niż gier atutowych wynikają prawdopodobnie z~faktu, że w~grach bez atutowych często łatwiej jest wyrobić i~wziąć na~forty (do~których gracze mogą dokładać dowolne z posiadanych kart), gdyż nie trzeba w~tym celu pozbawiać przeciwników atutów.


Poniższe twierdzenie uzasadnia poprawność metody szacowania wielkości drzewa użytej w~powyższym eksperymencie:
\begin{Tw}
Niech $D=(S, s_I, N)$ będzie drzewem o~skończonym zbiorze węzłów $S$, z~wyróżnionym korzeniem $s_I \in S$, o~funkcji następników $N: S \to 2^S$ ($N(s)$ to~zbiór dzieci węzła $s \in \S$). Przez $L$ oznaczmy zbiór liści drzewa $D$.

Rozważmy eksperyment polegający na przejściu drzewa, od~korzenia do~liścia poprzez losowe wybieranie kolejnych następników (z~jednostajnym rozkładem prawdopodobieństwa). Przez $P(l)$, gdzie $l \in L$, oznaczmy prawdopodobieństwo dojścia do~liścia $l$, tj.
	\[P(l) = \prod_{s \in W(l)} |N(s)|^{-1}\]
gdzie $W(l)$ to~zbiór wszystkich węzłów leżących na~ścieżce od $s_I$ do $l$, zawierający $s_I$ i~nie zawierający $l$.

Ponadto, z~powyższym eksperymentem zwiążemy zmienną losową $X: L \to \mathbb{N}$ zdefiniowaną, dla dowolnego $l \in L$, równością:
	\[X(l) = \prod_{s \in W(l)} |N(s)|\]
 
Wtedy:
	\[ EX = |L| \]
\end{Tw}
\begin{proof}
\[ EX = \sum_{l \in L} P(l)X(l) = \sum_{l \in L} 1 = |L|	\qedhere \]
\end{proof}

\bibliographystyle{plplain}
\bibliography{all}

\end{document}

%% file: lib_qwak.tex
\usepackage[utf8x]{inputenc}
\usepackage{polski}	
\usepackage{amsfonts}

\usepackage{caption}
\usepackage{cite}

\usepackage{tabularx}

\usepackage{tikz}
\usetikzlibrary{shapes, fit, backgrounds, shadows} 

\usepackage{pgfplots}
\pgfplotsset{compat=newest, /pgf/number format/use comma, /pgf/number format/1000 sep={}}	

\usepackage{import}

\usepackage{amsmath, amsthm, amssymb}

\theoremstyle{plain}
\newtheorem{Tw}{Twierdzenie}

\theoremstyle{definition}

\theoremstyle{remark}

\newcommand{\N}{\ensuremath{N}}

\renewcommand{\S}{\ensuremath{\mathbb{S}}}

\newcommand{\sym}[1]{\ensuremath{\textit{\text{#1}}}}	

\newcommand{\SUITS}{\sym{SUITS}}
\newcommand{\HANDS}{\sym{HANDS}}

\usepackage{listings}

\usepackage[breaklinks]{hyperref}
\hypersetup{colorlinks=true,linkcolor=black,citecolor=black,filecolor=black,urlcolor=black,breaklinks=true}

\newcolumntype{L}[1]{>{\raggedright\let\newline\\\arraybackslash\hspace{0pt}}m{#1}}
\newcolumntype{C}[1]{>{\centering\let\newline\\\arraybackslash\hspace{0pt}}m{#1}}
\newcolumntype{R}[1]{>{\raggedleft\let\newline\\\arraybackslash\hspace{0pt}}m{#1}}